\documentclass[aps,preprint,showpacs,groupedaddress,floatfix]{revtex4}
\usepackage{graphicx}
\usepackage{dcolumn}
\usepackage{bm}

\begin{document}

\title{Nuclear pairing from chiral pion-nucleon dynamics}
\author{N. Kaiser}
\affiliation{Physik-Department der Technischen Universit\"at M\"unchen, 
D-85748 Garching,
Germany}
\author{T. Nik\v si\' c}
\author{D. Vretenar}
\affiliation{Physics Department, Faculty of Science, University of Zagreb, 
Croatia}
\date{\today}

\begin{abstract}
We use a recently improved version of the chiral nucleon-nucleon potential at
next-to-next-to-leading order to calculate the $^1\!S_0$ pairing gap in
isospin-symmetric nuclear matter. The pairing potential consists of the 
long-range one- and two-pion exchange terms and two short-distance
NN-contact couplings. We find that the inclusion of the two-pion exchange at 
next-to-next-to-leading order reduces substantially the cut-off dependence of 
the $^1\!S_0$ pairing gap determined by solving a regularised BCS equation. 
Our results are close to those obtained with the universal low-momentum 
nucleon-nucleon potential $V_{\rm low-k}$ or the phenomenological Gogny D1S 
force.

\end{abstract}

\maketitle 

The self-consistent mean-field framework, extended to take into account 
the most important correlations, provides at present the only viable 
microscopic description of structure phenomena in light and 
heavy nuclei over the entire periodic table. 
A broad range of successful applications to 
nuclear structure and low-energy dynamics characterizes mean-field 
models based on the Gogny interaction, the Skyrme energy functional,
and the relativistic meson-exchange effective Lagrangian \cite{BHR.03,rmf}.
The effective forces used
in these models contain a moderate number of free parameters that are
adjusted to global properties of a small set of spherical and stable nuclei, 
rather than to the observables of free NN-scattering. In other words, 
even though the global effective nuclear interactions model the 
interaction between nucleons in the nuclear medium, they are 
not necessarily related to any given NN potential.
 
On the other hand, a completely new framework has recently been developed
in which nuclear interactions are formulated in terms of
effective field theories 
\cite{weinberg,kaplan,evgeni}. The key element is the separation of 
long- and short-distance dynamics and an ordering scheme in powers of small 
momenta. The nucleon-nucleon potential, constructed in the chiral perturbation 
theory, consists of the long-range contributions generated by one-, two- and 
three-pion exchange \cite{machleidt,evgenineu}, and a set of contact-terms 
encoding the short-distance NN-dynamics. The associated low-energy constants 
are adjusted to the empirical NN-phase shifts and deuteron properties. It has been
shown that their values can be understood in terms of the heavy-mass 
resonance exchanges \cite{reso}. Furthermore, when the isospin-breaking
corrections are systematically included, the chiral nucleon-nucleon potential 
reaches almost the same accuracy as the more phenomenological ``high-precision''
nucleon-nucleon potentials. This has been demonstrated in numerous
calculations of two- and few nucleon systems. 

Another line of approach, developed by the Stony-Brook group, applies
renormalisation group arguments to eliminate the high-momentum components from
phase-shift equivalent nucleon-nucleon potentials. By integrating out the
high-momentum components below a scale $\Lambda \approx 2$ fm$^{-1}$, a 
``universal'' 
low-momentum nucleon-nucleon $V_{\rm low-k}$ potential emerges \cite{vlowk}. 
Consequently, this potential operates only in
the subspace of nucleon states with momenta $p\leq \Lambda $.   
The pairing properties of nuclear matter derived from the $V_{\rm low-k}$ 
potential have been investigated recently in 
Ref. \cite{sedrakian}. Good agreement with the phenomenological Gogny pairing
interaction has been found for a wide range of nuclear densities. 
Calculations of nuclear matter properties have also been performed using the
$V_{\rm low-k}$ potential. 
The results obtained so far in the Hartree-Fock and Brueckner-Hartree-Fock 
approximations are however unsatisfactory because no saturation occurs in the 
nuclear matter equation of state \cite{kuckei}.
It has been concluded that the Brueckner-Hartree-Fock approximation, based on
the $V_{\rm low-k}$ potential, is applicable only at very low densities.
These findings are
consistent with the fact that nuclear pairing is known to be primarily
a low-density phenomenon. The maximum of the $^1\!S_0$ pairing gap typically 
lies at a density $\rho \approx  \rho_0/4$, where $\rho_0$ denotes the nuclear
matter saturation density. 
The purpose of the present paper is to investigate the pairing properties of the
chiral nucleon-nucleon potential.  
            
Quite generally, the momentum and density-dependent pairing field 
$\Delta(k,k_f)$ in infinite nuclear matter is determined by the solution of 
the BCS gap equation
\begin{equation}
\label{gapeq}
\Delta(k,k_f) = -\frac{1}{4\pi^2}
  \int_0^\infty{\frac{p^2 V(p,k) \Delta(p,k_f)}
   {\sqrt{[ {\cal E}(p,k_f)-{\cal E}(k_f,k_f)]^2+\Delta(p,k_f)^2}}\;dp} \;.
\end{equation}     
$V(p,k)$ represents the off-shell pairing potential in momentum space,
${\cal E}(p,k_f)$ is the quasiparticle energy, and ${\cal E}(k_f,k_f)$ is the
Fermi energy. 

The effective force in the pairing channel is generated by the sum
of all particle-particle
irreducible Feynman diagrams~\cite{LM.63,BW.63,Mig.67}. In most application to
nuclear and neutron matter, however, 
only the lowest-order term, which corresponds to the bare nucleon-nucleon 
interaction, is retained~\cite{DH-J.03}. Terms of higher order in the effective
pairing interaction represent screening corrections to the bare force, caused by
medium polarization effects. Numerous studies have
shown that polarization effects can have a pronounced
influence on the calculated values of the pairing
gaps (see, for instance, Refs.~\cite{Cla.76,AWP.89,Sch.96}). 
The influence of both the vertex corrections to the pairing interaction, and 
the self-energy corrections, on the properties of $^1S_0$ pairing in 
neutron and nuclear matter has recently been
studied in the framework of the generalized gap equation~\cite{SLS.03,LSS.04}.
It has been found that the two effects lead to a strong suppression of the
pairing correlations in neutron matter (the pairing gap is reduced by 
more than 50\% with respect to the BCS prediction), whereas 
they cancel each other out to a large extent in symmetric nuclear matter.
The pairing correlation energy in 
finite nuclei can be calculated in the local density approximation (LDA).
LDA calculations with the Gogny D1 force have been compared 
with exact Hartree-Fock-Bogoliubov (HFB) calculations of the pairing
correlation energy for many spherical nuclei \cite{Kuc.89}.  
Except for shell effects, the results of these calculations are in 
close agreement, and thus one should not expect large effects from 
medium corrections not included in the BCS limit.
In the present analysis we only consider symmetric nuclear matter and, 
therefore, medium polarization effects will not be taken into account
at this stage.

For the pairing potential  $V(p,k)$ we employ the improved version of the 
chiral nucleon-nucleon potential derived in Refs. \cite{EGM.03,EGM.04}. An
overly strong medium-range attraction in the isoscalar central part of the
chiral two-pion exchange at the next-to-next-to-leading order in the chiral 
expansion, present in earlier versions of the chiral potential \cite{evgeni}, 
has been removed by using the spectral function regularization method. 
Essentially, this means that only $\pi\pi$-intermediate states of invariant
mass below a scale $\tilde \Lambda$ (where the chiral effective field theory is
applicable) are taken into account in the pion-loop integrals, while shorter
range contributions should be represented by NN-contact interactions.

The bare chiral NN-potential used in the present study consists of the 
one-pion exchange term, the NN-contact interaction, and the irreducible 
two-pion exchange term
\begin{equation}
\label{ChPTpot}
V(p,k)=V^{(1\pi)}(p,k) + V^{(ct)}(p,k)+V^{(2\pi)}(p,k)\;.
\end{equation}
This approximation for the two-nucleon potential is valid for small values of 
the momenta $p$ and $k$ and it breaks down for momenta above the chiral
symmetry breaking scale. An additional cut-off  $\Lambda$, included by 
multiplying the  potential $V(p,k)$ with a regulator function $f^\Lambda(p)$
\cite{evgeni}  
\begin{equation}
\label{regulator}
V(p,k) \to f^\Lambda (p)V(p,k)f^\Lambda (k)\;,
\end{equation}
prevents the growth of the potential with increasing momenta $p$ and $k$.
Following the procedure of Ref. \cite{EGM.04}, we employ the exponential
regulator function
\begin{equation}
\label{regulator_fun}
f^\Lambda (p)=\exp(-p^6/\Lambda^6) \;.
\end{equation}
Even though both cut-off parameters: $\tilde{\Lambda}$ and $\Lambda$, 
are introduced to remove high momentum components of the interacting 
nucleon and pion fields, their roles are different. The inclusion of 
$\tilde{\Lambda}$ removes the short-distance portion of the TPE 
component of the nuclear force, whereas the parameter
$\Lambda$ ensures that high-momentum nucleon states do not contribute to the
scattering process. If all terms in the EFT expansion are included, 
low-energy observables should not depend on the cut-off parameters. 
In practice, however, the expansion is always truncated at some order. 
Consequently, the observables depend on the cut-off parameters to some extent, 
but this dependence becomes weaker when higher-order terms are included. 
Unless stated otherwise, we use the value $\Lambda=550$ MeV in 
the following calculations. 

The one-pion exchange contribution in the S-wave channel reads
\begin{equation} 
\label{onepi}
V^{(1\pi)}(p,k) = \frac{g_A^2}{2f_\pi^2} \bigg\{ 1 - {m_\pi^2 \over 4 p k}
\ln{\frac{m_\pi^2+(p+k)^2}{m_\pi^2+(p-k)^2}} \bigg\} \;.
\end{equation} 
The following numerical values have been used for the
nucleon axial vector coupling constant $g_A$, the weak pion decay constant
$f_\pi$, and the pion mass $m_\pi$: $g_A = 1.3$, $f_\pi =92.4$ MeV, $m_\pi=135$
MeV.

The NN-contact interactions includes the contact operators without and with two
derivatives
\begin{equation} 
\label{contact}
V^{(ct)}(p,k) =  \frac{\widetilde{C}_{^1\!S_0}}{2\pi} + 
\frac{{C}_{^1\!S_0}}{2\pi}(p^2+k^2)\;. 
\end{equation}
For each choice of the cut-off parameters $\tilde{\Lambda}$ and $\Lambda$, 
the low-energy constants $\widetilde{C}_{^1\!S_0}$ and $C_{^1\!S_0}$ are 
determined by a fit to the $^1\!S_0$ phase shifts below the inelastic
$NN\pi$-threshold. The cut-off dependence entering through the regulator 
function $f_\Lambda(p)$ is to a large extent cancelled by that of the running 
low-energy constants $\widetilde{C}_{^1\!S_0}$ and $C_{^1\!S_0}$. In the 
present calculation we employ the central values of the low-energy constants 
at the next-to-leading order (NLO) and next-to-next-to-leading order (NNLO) 
listed in Table 4 of Ref. \cite{EGM.04}. These particular values of the 
low-energy constants have been adjusted for $\tilde{\Lambda}=600$ MeV and 
$\Lambda=550$ MeV.
 
The irreducible two-pion exchange contribution is most conveniently calculated
via a twice subtracted dispersion relation 
\begin{eqnarray} 
\label{twopi}
V^{(2\pi)}(p,k) &=& \frac{1}{\pi} \int_{2m_\pi}^{\tilde{\Lambda}} d\mu 
\, {\rm Im}[V_C + W_C -2\mu^2 V_T - 2\mu^2 W_T]
\nonumber \\ && 
\times \bigg\{ \frac{4}{\mu} - \frac{4}{\mu^3}(p^2+k^2) -
\frac{\mu}{pk} \ln{\frac{\mu^2+(p+k)^2 }{\mu^2+(p-k)^2}} \bigg\}  \;.
\end{eqnarray}
Im$V_C$, Im$W_C$, Im$V_T$, and Im$W_T$ are the imaginary parts of the 
isoscalar central, isovector central, isoscalar tensor, and isovector tensor 
NN-amplitudes. The weighting function in curly brackets 
originates from projecting the NN-potential onto the spin-singlet S-wave. 
The cut-off parameter $\tilde{\Lambda}$, which restricts the spectral functions
calculated in the chiral perturbation theory to their low-energy domain of
validity, is kept fixed throughout this work at the value $\tilde{\Lambda}
=600$ MeV. The imaginary parts of the isovector central and isoscalar 
tensor NN-amplitudes contribute at the NLO, while the 
imaginary parts of the isoscalar central and isovector tensor 
NN-amplitudes contribute at the NNLO in the chiral
expansion. Explicitly, the spectral functions corresponding to the two-pion
exchange at the NLO read 
\begin{equation} 
\label{WC}
{\rm Im}W_C(i\mu) = \frac{\sqrt{\mu^2-4m_\pi^2}}{3\pi \mu (4f_\pi)^4} 
\bigg[ 4m_\pi^2(1+4g_A^2-5g_A^4) +\mu^2(23g_A^4-10g_A^2-1) + 
{48 g_A^4 m_\pi^4 \over \mu^2-4m_\pi^2} \bigg] \;, 
\end{equation}
\begin{equation} 
\label{VT}
{\rm Im}V_T(i\mu) = - 
\frac{6 g_A^4 \sqrt{\mu^2-4m_\pi^2}}{\pi  \mu (4f_\pi)^4}\;, 
\end{equation}
and at the NNLO  
\begin{equation} 
\label{VC}
{\rm Im}V_C(i\mu) = \frac{3g_A^2}{64 \mu f_\pi^4} \Big[
(4c_1-2c_3)m_\pi^2 + c_3\, \mu^2 \Big] (2m_\pi^2-\mu^2) \;, 
\end{equation} 
\begin{equation} 
\label{WT}
{\rm Im}W_T(i\mu) = \frac{g_A^2 c_4}{128 \mu f_\pi^4} (4m_\pi^2-\mu^2) \;, 
\end{equation}  
with the low-energy constants: $c_1 = -0.81\,$GeV$^{-1}$, $c_3=-3.40\,
$GeV$^{-1}$, and $c_4 =3.40\,$GeV$^{-1}$ \cite{EGM.03,EGM.04,nnpap}.

For the single-particle spectrum which enters the gap equation (\ref{gapeq}), 
we employ the simple quadratic form \begin{equation}
\label{spectrum} 
{\cal E}(p,k_f)- {\cal E}(k_f,k_f)= \frac{p^2-k_f^2}{2 M^*(k_f)} \;.
 \end{equation}
Such an approximation is sufficient since momenta $p$ around $k_f$ give the
dominate contribution to the integral in Eq. (\ref{gapeq}).
The effective nucleon mass $M^*(k_f)$ is deduced from a very recent three-loop 
calculation of nuclear matter in chiral perturbation theory including explicit
$\Delta(1232)$-isobar degrees of freedom \cite{FKW.04}. 
 
\begin{figure}
\begin{center}
\includegraphics[scale= 0.6,clip]{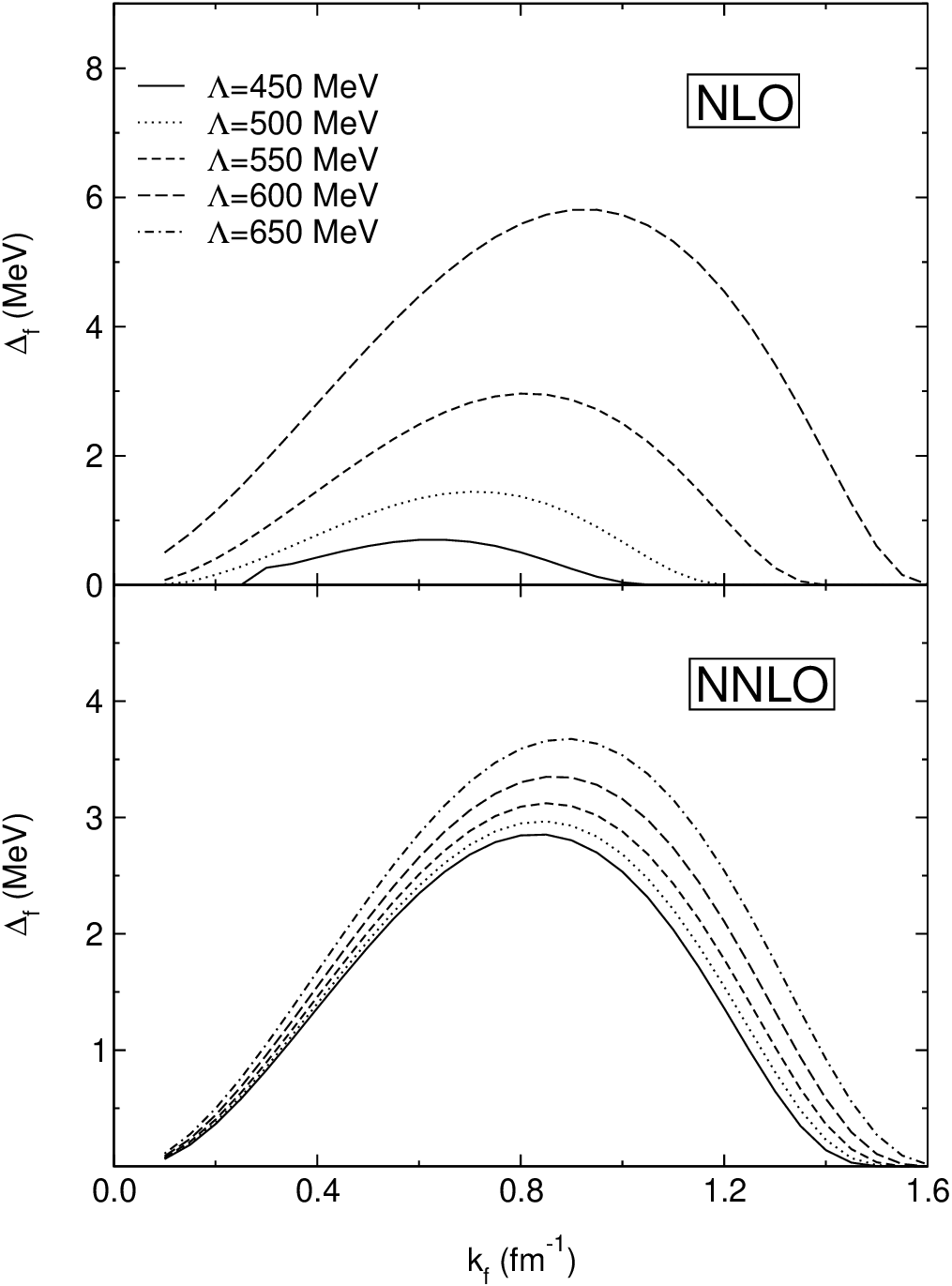}
\end{center}
\caption{Pairing gap at the Fermi surface $\Delta_f \equiv \Delta(k_f,k_f)$
as function of the Fermi momentum, for different values of the 
cut-off parameter $\Lambda$, at NLO (upper panel) and NNLO (lower panel).}
\label{FigA}
\end{figure}

As we have already emphasized, the low-energy observables should not depend 
sensitively on the precise value of the cut-off parameter $\Lambda$.
In addition, the dependence should become weaker when higher order terms are
included in the small momentum expansion. To verify this we have calculated the
pairing gaps $\Delta_f \equiv \Delta(k_f,k_f)$ in symmetric nuclear matter
for several values of the cut-off parameter $\Lambda$ between $450$ MeV and 
$650$ MeV.
For the low-energy constants $\widetilde{C}_{^1\!S_0}$ and ${C}_{^1\!S_0}$ we 
have used the central values listed in Table 4 of Ref.~\cite{EGM.04}, and the
spectral function cut-off $\tilde{\Lambda}$ has been fixed at $\tilde
{\Lambda}=600\;$MeV. 
Notice that no attempt has been made to readjust the LECs when 
changing the cut-off parameter $\Lambda$. 
The results are displayed in Fig.~\ref{FigA}. At the NLO the pairing gap 
strongly depends on the value of $\Lambda$. With increasing the 
cut-off from $\Lambda=450$ MeV to $\Lambda=600$ MeV, the maximum value of
the pairing gap increases from $\Delta_f^{max}=0.7$ MeV to 
$\Delta_f^{max}=5.81$ MeV, and the value of the Fermi momentum for which 
the pairing gap reaches maximum changes from 
$p_f^{max}=0.65$ fm$^{-1}$ to $p_f^{max}=0.95$ fm$^{-1}$.
For $\Lambda=650$ MeV the maximum value of the gap is well above 
10 MeV and the calculation obviously diverges. The results are much
improved at the NNLO: the position of the maximum changes very little
with increasing $\Lambda$, and the value of the pairing gap at 
maximum increases less than $0.8$ MeV between 
$\Lambda=450$ MeV and $\Lambda=650$ MeV. Since the gap at the Fermi 
surface is very sensitive to changes in the effective pairing potential,
this indicates rapid convergence for the EFT expansion.

\begin{figure}
\begin{center}
\includegraphics[scale= 0.6,clip]{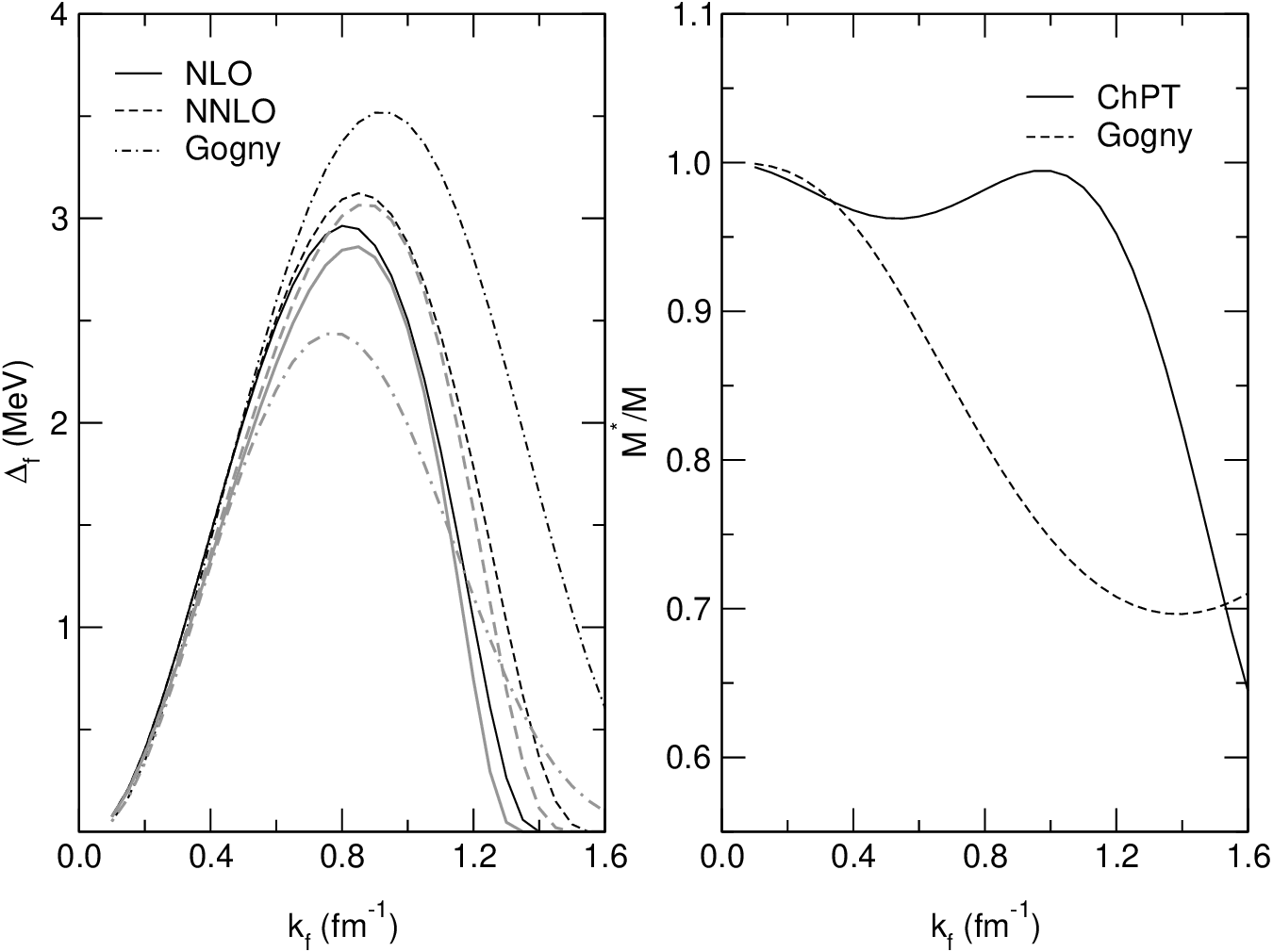}
\end{center}
\caption{Pairing gaps at the Fermi surface $\Delta_f = \Delta(k_f,k_f)$
in symmetric nuclear matter, calculated with the chiral NN-potential 
(\protect\ref{ChPTpot}) at NLO and NNLO, and with the effective Gogny 
interaction D1S~\protect\cite{BGG.91}. The heavy and light curves refer to 
calculations with non-interacting and medium-modified single-particle spectra,
respectively. The panel on the right illustrates the density dependence of 
the effective nucleon masses for the Gogny interaction and for the chiral
potential \protect\cite{FKW.04}. }
\label{FigB}
\end{figure}

In the left panel of Fig.~\ref{FigB} we compare the density dependence
of the pairing gaps in nuclear matter, calculated by employing 
the chiral NN-potential and with the effective Gogny
interaction. The solid and the dashed curves correspond to the
chiral potential at the NLO and the NNLO, respectively. The dash-dotted curve
is the pairing gap calculated with the D1S Gogny interaction~\cite{BGG.91}.
Similar to the analysis of Ref.~\cite{sedrakian}, where the Gogny pairing gaps 
were compared with those calculated with the low-momentum $V_{{\rm low}-k}$
interaction extracted from realistic forces, we have calculated the 
pairing gaps using both the non-interacting
single-particle spectra (heavy curves), and the medium-modified single-particle
spectra (light curves). The medium-modified
single-particle spectra are computed in the
Hartree-Fock approximation for the Gogny interaction, and by using
Eq.~(\ref{spectrum}) for the chiral NN-potential. 
First we notice that for the noninteracting single-particle spectra
there is a qualitative and, at low densities $k_f \leq 0.8$ fm$^{-1}$,
even a quantitative agreement between the pairing gaps 
calculated with the chiral NN potential and the effective Gogny interaction.
The inclusion of the NNLO contribution increases the value of the 
pairing gap at maximum and shifts the position of the maximum to 
a slightly higher value of $k_f$, in closer agreement with the 
Gogny pairing gap. For $k_f > 1$ fm$^{-1}$ and closer to saturation 
density, we find a more pronounced difference between the 
Gogny pairing gap and the gaps calculated with the chiral NN
potential.  This difference with respect to the Gogny pairing 
gap at higher densities has also been observed for other 
bare NN interactions \cite{Garr.99,Serra.01,sedrakian}. 

The effective nucleon mass is reduced in the nuclear medium. 
This results in a lower
density of states around the Fermi surface and, hence, 
in weaker pairing correlations. The reduction of the pairing gap 
is much more pronounced for the Gogny effective interaction than 
for the chiral NN potential. This is caused by a much stronger 
reduction of the effective mass for the former interaction in 
the relevant range of densities $0.4 \leq k_f \leq 1.2$ fm$^{-1}$
The effective nucleon mass $M^*(k_f)/M$ from the chiral 
EFT three-loop calculation \cite{FKW.04}, and for 
the D1S Gogny effective interaction are shown in 
the right panel of Fig.~\ref{FigB}. The density dependence is very 
different in the two cases, and this means that one should not, at least
on the quantitative level, compare the corresponding pairing gaps 
calculated with medium-modified single-nucleon spectra.

\begin{figure}
\begin{center}
\includegraphics[scale= 0.6,clip]{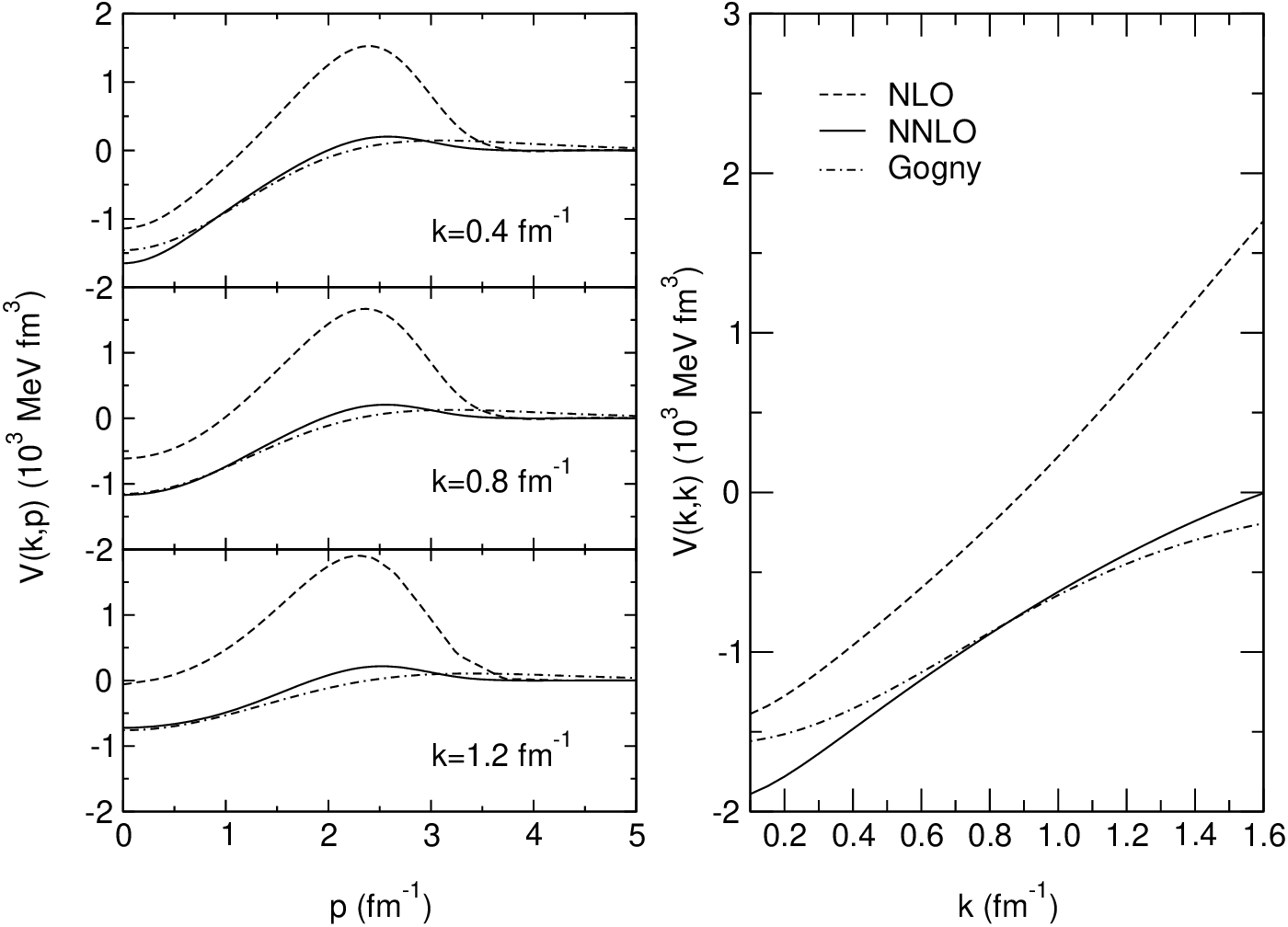}
\end{center}
\caption{Momentum dependence of the pairing potentials $V(p,k)$: chiral 
NN-potential at NLO and NNLO, and the Gogny DS1 interaction, for three 
different values of the  momenta $k=0.4, 0.8, 1.2$ fm$^{-1}$.
In the panel on the right the corresponding diagonal matrix elements
$V(k,k)$ are shown as functions of the momentum $k$.}
\label{FigC}
\end{figure}

In Fig.~\ref{FigC} we plot the momentum dependence of the 
pairing potentials $V(p,k)$ for three different values of the 
momenta $k=0.4, 0.8, 1.2$ fm$^{-1}$. While the Gogny 
interaction and the chiral NN-potential at the NLO display only a
qualitatively similar momentum dependence, the inclusion of the NNLO
contributions brings the chiral NN-potential in 
remarkable  agreement with the Gogny D1S interaction. The 
same  is true and even more pronounced, of course, for the diagonal 
matrix elements  $V(k,k)$ shown in the right panel of Fig.~\ref{FigC}.
A similar result has been obtained for the  $V_{\rm low-k}$ potential
in Ref.~\cite{sedrakian}.

In conclusion, we have presented a study of the chiral nucleon-nucleon
potential pairing properties. The potential consists of the long-range one-
and two-pion exchange terms, and it includes 
the short-distance NN-contact terms.
Our main result is that the inclusion of the two-pion exchange 
at the next-to-next-to-leading order reduces substantially the cut-off 
dependence of the $^1\!S_0$ pairing gap, indicating good convergence of the
small momentum expansion. In addition, 
the inclusion of the NNLO contributions brings
the chiral NN-potential in remarkable  agreement with the Gogny D1S interaction
and the $V_{\rm low-k}$ potential.
In future we plan to include the NNNLO terms in the chiral expansion of the
bare NN-potential and extend our study to the neutron matter pairing
properties. This will necessitate the inclusion of medium modification 
effects, such as the polarization effects, vertex corrections and 
Pauli-blocking effects. The chiral effective field theory at finite density
represents a systematic framework for such extension. Work along these lines
is in progress.

\leftline{\bf ACKNOWLEDGMENTS}
We thank P. Ring for useful discussions and 
E. Epelbaum and S. Fritsch for supplying numerical results.

\newpage

\end{document}